\documentclass[fleqn,usenatbib]{mnras}

\usepackage{newtxtext,newtxmath}

\usepackage[T1]{fontenc}
\usepackage{times}

\usepackage{graphicx}	
\usepackage{amsmath}	
\usepackage{amssymb}	
\usepackage{mathtools}  
\usepackage{microtype}  
\usepackage{ragged2e}
\usepackage{xparse}     

\newcommand{\Rv}{{\ensuremath\text{R}_{200}}}
\newcommand{\Mv}{{\ensuremath\text{M}_{200}}}
\newcommand{\cv}{{\ensuremath\text{c}_{200}}}
\newcommand{\Msun}{{\ensuremath\text{M}_\odot}}
\newcommand{\specline}[2]{\ensuremath{\mathchoice{{\text{#1}}\,{#2}}
                                                 {{\text{#1}}\,{#2}}
                                                 {{\text{#1}}{#2}}
                                                 {{\text{#1}}{#2}}}}
\newcommand{\Ha}{{\specline{H}{\alpha}}}

\newcommand{\Hmol}{{\ensuremath\text{H}_2}}
\newcommand{\spec}[2]{{\text{#1}\,\textsc{#2}}}
\newcommand{\HI}{\spec{H}{i}}
\newcommand{\HII}{\spec{H}{ii}}
\newcommand{\HeI}{\spec{He}{i}}
\newcommand{\HeII}{\spec{He}{ii}}
\newcommand{\HeIII}{\spec{He}{iii}}

\newcommand{\sbunit}{{\ensuremath\text{erg}\,\text{s}^{-1}\text{cm}^{-2}\text{arcsec}^{-2}}}
\newcommand{\aunit}{{\text{\AA}}}
\newcommand{\pcm}{{\ensuremath\text{cm}^{-2}}}

\newcommand{\primHe}{{\ensuremath y_{\text{P}}}}
\newcommand{\primHeMass}{{\ensuremath Y_{\text{P}}}}
\newcommand{\emr}{{\ensuremath R_{\varepsilon}}}
\newcommand{\sbr}{{\ensuremath R_{\Sigma}}}

\newcommand{\lr}{{\ensuremath R_{\text{F}}}}
\newcommand{\her}{{\ensuremath R_{\text{He}}}}
\newcommand{\lerr}{{\ensuremath\delta_{\text{F}}}}
\newcommand{\aerr}{{\ensuremath\delta_{\aUV}}}
\newcommand{\yperr}{{\ensuremath\delta_{\primHe}}}
\newcommand{\aUV}{{\ensuremath\alpha_{\text{UV}}}}
\newcommand{\ICF}{{\text{ICF}}}

\NewDocumentCommand \df {m O{d} o}{{\ensuremath\text{#2}\IfValueT{#3}{^{#3}}#1}}
\NewDocumentCommand \dff {G{} m o}{{\ensuremath\frac{\df{#1}[#3]}{\df{#2}[#3]}}}

\defcitealias{benitez-llambayPropertiesDarkLCDM2017}{BL17}
\defcitealias{sykesFluorescentRingsStarfree2019}{S19}
\defcitealias{madauCosmicReionizationPlanck2015}{MH15}

\newcommand{\graphiccommand}[1]{\includegraphics{embed-#1-crop}}

\title[$Y_{\text{P}}$ and the UVB slope from fluorescent RELHICs]{Determining the primordial helium abundance and UV background using fluorescent emission in star-free dark matter haloes}

\author[C. Sykes et al.]{
Calvin Sykes$^{1,2}$\thanks{E-mail: calvin.v.sykes@durham.ac.uk},
Michele Fumagalli$^{1,2,3}$,
Ryan Cooke$^{2}$,
Tom Theuns$^{1}$
\\
$^{1}$Institute for Computational Cosmology, Durham University, Durham DH1 3LE, UK\\
$^{2}$Centre for Extragalactic Astronomy, Durham University, Durham DH1 3LE, UK\\
$^{3}$Dipartimento di Fisica G. Occhialini, Universit\`a degli Studi di Milano Bicocca, Piazza della Scienza 3, 20126 Milano, Italy
}

\date{Accepted XXX. Received YYY; in original form ZZZ}

\pubyear{2019}

\begin{document}
\label{firstpage}
\pagerange{\pageref{firstpage}--\pageref{lastpage}}
\maketitle

\begin{abstract}
Observational measures of the primordial helium mass fraction, $\primHeMass$, are of interest for cosmology and fundamental particle physics.
Current measures obtained from $\HII$ regions agree with the Standard Model prediction to approximately 1\% precision, although these determinations may be affected by systematic uncertainties.
This possibility can only be tested by independently measuring the helium abundance in new ways.
Here, we propose a novel method to obtain a measurement of $\primHeMass$ using hydrogen and helium recombination line emission from RELHICs: pristine, gas-rich but star-free low-mass dark matter haloes whose existence is predicted by hydrodynamical simulations.
Although expected to be uncommon and intrinsically faint in emission, the primordial composition and simple physical properties of these objects make them an ideal laboratory to determine $\primHeMass$.
We present radiative transfer simulations to demonstrate the effectiveness of this approach, finding that comparing the emission in H and He lines, either via their volumetric emissivities, or integrated properties such as the surface brightness and total flux, may be used to infer $\primHeMass$.
Furthermore, we show that RELHICs can be used to provide an entirely novel constraint on the spectral slope of the ultraviolet background, and discuss the possibility of measuring this slope and the primordial helium abundance simultaneously.
\end{abstract}




\section{Introduction}
\label{sec:intro}

Almost all helium atoms in the Universe were synthesised in the first few minutes after the Big Bang, during the period of Big Bang nucleosynthesis~\citep[BBN;][]{alpherOriginChemicalElements1948,hoyleMysteryCosmicHelium1964}.
The primordial helium mass fraction $\primHeMass$, or equivalently the abundance by number $\primHe$\footnote{These two quantities are related by $\primHeMass=4\primHe/(1+4\primHe)$. We note that $\primHeMass$ is defined as $\primHeMass\equiv4\,n(^{4}{\text{He}})/n_{\text{b}}$, where $n_{\text{b}}$ is the baryon density. $\primHeMass$ is therefore somewhat of a misnomer; it does not represent the mass fraction of ${}^{4}\text{He}$. Since BBN codes naturally calculate a number abundance ratio, and observations also measure the primordial helium abundance in this form, we will predominantly use the number abundance $\primHe\equiv n_{\text{He}}/n_{\text{H}}$ in this paper.}, which results from this brief period of nucleosynthesis is influenced by the early-time expansion history of the Universe, and by the abundance of free neutrons at the onset of nucleosynthesis, which in turn depends on the neutron half-life.
The primordial helium abundance is therefore sensitive to both cosmology and particle physics, making accurate measurements of this quantity highly informative.
The precise measurements of the baryon-to-photon ratio obtained from cosmic microwave background satellites such as \emph{Planck} remove the final free parameter from BBN, meaning that the Standard Model prediction of $\primHeMass=0.24672 \pm 0.00017$ ($\primHe=0.08188\pm0.00008$) is reliable~\citep{pitrouPrecisionBigBang2018}.

Comparing observational measures of $\primHeMass$ with the BBN prediction allows the presence of any new physics beyond the Standard Model to be investigated.
To date, the leading method for determining $\primHeMass$ involves comparing the relative intensity of hydrogen and helium emission lines measured in $\HII$ regions, ionized bubbles of gas surrounding regions of active star formation~\citep[e.g.][]{izotovNewDeterminationPrimordial2014,averEffectsHeUplambda2015,peimbertPrimordialHeliumAbundance2016,valerdiDeterminationPrimordialHelium2019,fernandezPrimordialHeliumAbundance2018,fernandezBayesianDirectMethod2019}.
These studies select $\HII$ regions in metal-poor galaxies ($Z/Z_\odot \lesssim 0.1$), to minimise the enrichment by stars where emission is detected.
However, the level of contamination remains necessarily non-zero, so the observed ratios of hydrogen to helium emission must be extrapolated down to zero metallicity in order to recover the primordial abundance ratio.
This limitation introduces the possibility of systematic errors~\citep[see e.g.][]{izotovPrimordialAbundanceOf4He2007,porterUncertaintiesTheoreticalHe2009}, the characterisation of which becomes increasingly important as statistical errors on the measurements improve.
Consequently, it is beneficial to consider independent techniques for determining $\primHeMass$.
One such alternative involves studying intergalactic absorption lines arising in almost-primordial clouds located between us and a background quasar.
This approach has been demonstrated to yield a primordial value of $\primHeMass=0.250_{-0.025}^{+0.033}$ \citep{cookeMeasurementPrimordialHelium2018a} consistent with the Standard Model prediction, although the constraint obtained is not yet as tight as that resulting from $\HII$ region measurements, for which a weighted average of recent determinations (see references above) gives $\primHeMass=0.248\pm0.001$.

In this paper, we discuss a novel method for determining $\primHeMass$.
We focus on low-mass dark matter haloes, the existence of which is a robust prediction of the cold dark matter (CDM) model for hierarchical structure formation.
Below a mass scale of approximately $10^{10}\,\Msun$, observational constraints indicate that many haloes fail to host luminous galaxies~\citep{klypinWhereAreMissing1999,mooreDarkMatterSubstructure1999}.
This requirement can be met by appealing to baryonic feedback processes; most prominently, cosmic reionization heats intergalactic gas to ${\sim}10^4\,\text{K}$, inhibiting star formation in haloes with potential wells too shallow to confine the heated gas~\citep{miralda-escudeReionizationThermalEvolution1994,okamotoMassLossGalaxies2008,meiksinPhysicsIntergalacticMedium2009}.
Using the \textsc{Apostle} suite of Local Group hydrodynamical simulations~\citep{sawalaAPOSTLESimulationsSolutions2016}, \citet[hereafter~\citetalias{benitez-llambayPropertiesDarkLCDM2017}]{benitez-llambayPropertiesDarkLCDM2017} identified a population of haloes with masses $10^8<M_{\text{halo}}/\Msun<10^{9.6}$, which additionally experience negligible star formation prior to reionization.
Hence, these haloes remain essentially star-free down to redshift $z=0$, and by avoiding mechanisms such as ram pressure stripping from interactions with the cosmic web, can retain a small reservoir of essentially-pristine gas.
This gas consists of an approximately kiloparsec-sized neutral core surrounded by an envelope kept ionized by the diffuse ultraviolet background (UVB), motivating the naming of this population as ``REionization-Limited HI Clouds'' (RELHICs).

In a previous paper~\citep[hereafter~\citetalias{sykesFluorescentRingsStarfree2019}]{sykesFluorescentRingsStarfree2019}, we performed radiative transfer simulations to model RELHICs and examine the effects of the UVB on the properties of their gas.
The UVB ionizes atoms in the gas, which later recombine to produce hydrogen emission lines such as $\Ha$.
We found that for RELHICs with masses in the narrow range $10^{9.4}<M_{\text{halo}}/\Msun<10^{9.6}$, this fluorescent emission displays a distinctive ring-shaped morphology when seen in projection on the sky.
The narrow mass range for which we predict these fluorescent rings, in combination with their intrinsically low surface brightness of the emission, makes fluorescent rings rare and their detection challenging.
This intrinsic brightness increases at higher $z$, due to the greater amplitude of the UVB.
However, the need to resolve the ring-shaped emission that distinguishes a fluorescent RELHIC, in combination with the rapid onset of cosmological surface brightness dimming, means that only relatively local RELHICs ($z\lesssim 0.2$) are realistic candidates for detection.
Nevertheless, they remain a firm prediction of the CDM paradigm, and their detection would provide a probe of this cosmological model on an as-yet untested scale.
Furthermore, we have shown that observable properties of the rings, such as their projected size and peak brightness, are sensitive to the properties of the UVB and the mass of the underlying dark matter halo.

Fluorescent RELHICs will also produce emission in helium recombination lines, which will exhibit a similar ring-like appearance.
As a result of their star-free nature, the gas they contain should be almost pristine in composition, and so they have the potential to yield a direct constraint on $\primHe$, albeit one with substantial observational challenges given current instrumentation, as we will show.
In this paper, we explore this possibility, and find that in addition to being able to measure the helium abundance, a comparison of the fluorescent hydrogen and helium emission lines from RELHICs could provide the first observational constraint on the shape of the ionizing UVB spectrum.

The paper is organised as follows: in Section~\ref{sec:model}, we provide a brief description of our numerical method and describe how we have extended the calculations in \citetalias{sykesFluorescentRingsStarfree2019} to additionally predict surface brightnesses for $\HeI$ and $\HeII$ emission lines. We then present our results in Section~\ref{sec:results}, considering constraints on the primordial helium abundance $\primHe$ and on the UVB slope in turn (Sections~\ref{ssec:yp} and \ref{ssec:aUV}), and then combined constraints on both parameters (Section~\ref{ssec:combined}). We conclude by discussing our results and their implications in Section~\ref{sec:discuss}.
Throughout, we assume a set of cosmological parameters ($H_0=67.3\;\text{km}\,\text{s}^{-1}\,\text{Mpc}^{-1}$, $\Omega_\Lambda=0.685$, $\Omega_{\text{M}}=0.315$, $\Omega_{\text{B}}=0.0491$) consistent with \emph{Planck} measurements \citep{planckcollaborationPlanck2013Results2014}.

\section{Modelling helium emissivities}
\label{sec:model}

As was demonstrated in \citetalias{benitez-llambayPropertiesDarkLCDM2017}, the RELHICs identified in \textsc{Apostle} are well-described by a simple analytic model in which the gas they contain is in hydrostatic equilibrium with a gravitational potential due to the host dark matter halo, and in thermal equilibrium with the UVB.
To predict their emission properties, we implement this analytic model using an ionization balance code originally described in \citet{cookePrimordialAbundanceDeuterium2016} and with additional modifications introduced by \citetalias{sykesFluorescentRingsStarfree2019}.\footnote{\RaggedRight The code is made available at: \url{https://github.com/calvin-sykes/spherical_cloudy}}
Similar in functionality to photoionization codes such as \textsc{cloudy} \citep{ferland2017ReleaseCloudy2017a}, our code includes two important additional features.
Firstly, it applies the condition of hydrostatic equilibrium to determine the gas density profile, using the background gravitational of a Navarro-Frenk-White \citep[NFW;][]{navarroStructureColdDark1996} dark matter halo.
Secondly, although we assume spherical symmetry (implying that the UVB irradiates the gas isotropically), the gas column density has an angular dependence which also varies with radius, meaning that the local attenuated radiation field, and hence the local photoionization and photoheating rates, are functions of both depth within the cloud and direction of incident radiation.
Consequently, we perform our calculations in (projected) spherical coordinates.

Our calculations take the following form, which is similar to that described by \citet{sternbergAtomicHydrogenGas2002}.
A dark matter potential is defined by choosing a virial mass $\Mv$ and obtaining the concentration parameter $\cv \equiv \Rv/r_s$ from the \citet{ludlowMassConcentrationRedshift2016} mass-concentration relation.
Here, $r_s$ is the NFW scale radius, and virial quantities are defined such that within a sphere of radius $\Rv$, the average density is $200\rho_{\rm{crit}}$, where $\rho_{\rm{crit}}$ is the critical density of the Universe, and the total mass enclosed by this sphere is $\Mv$.
A total baryonic gas mass $\text{M}_{\text{g}}$ is then assigned to the halo using the analytic model employed by \citetalias{benitez-llambayPropertiesDarkLCDM2017}; the gas is split into $N_r=1000$ radial cells and initialised to be fully ionized and isothermal with temperature $T=10^4\,$K.
We assume the gas to have primordial composition, with the helium abundance given by $n_{\text{He}}=\primHe n_{\text{H}}$.

We then determine the pressure profile required for hydrostatic equilibrium, using as a boundary condition the assumption that at $r\gg\Rv$, the gas density approaches the cosmic mean baryon number density, $\bar{n}_{\text{H}} \simeq 10^{-6.7} \rm{cm}^{-3}$ at $z=0$.
From the resulting gas density profile, we next determine the intensity of the radiation field within the cloud, using the \citet[][hereafter \citetalias{madauCosmicReionizationPlanck2015}]{madauCosmicReionizationPlanck2015} $z=0$ UVB as the initial, unattenuated spectrum. We calculate ionization rates for photoionization, primary and secondary collisional ionizations, hydrogen ionization resulting from helium recombination radiation, and charge transfer ionization.
We further calculate recombination rates for radiative, dielectronic and charge transfer processes, assuming Case B conditions hold throughout the cloud.
This means we ignore recombinations directly to the ground state, because at the typical densities associated with the ionization front, the ionizing photons produced by these recombinations will invariably ionize a nearby neutral atom.
Thus, they will have no effect on the overall ionization state of the gas and produce no detectable emission.\footnote{We further justify this assumption in Appendix B of \citetalias{sykesFluorescentRingsStarfree2019}.}
By equating the relevant processes for each atomic species, we enforce ionization equilibrium and thus determine the fractional ionization $X_{A^{i+}} \equiv n_{A^{i+}} / n_A$ of all species $A$ (here H and He) and ionization stages $i$.

Finally, we determine the temperature profile by assuming thermal equilibrium for higher-density gas ($n_{\text{H}}> 10^{-4.8}\,\text{cm}^{-3}$), where this threshold is set by the condition that the timescale for equilibrium must be shorter than the Hubble time. For gas below this threshold density, we instead set the gas temperature to that resulting from a heating timescale equal to the Hubble time, interpolating between the two regimes to ensure the temperature profile remains smooth.
In the equilibrium case, we compute the heating rate by considering primary photoheating and secondary heating by primary photoelectrons, while the cooling rate includes contributions from collisional excitation/ionization cooling, recombination cooling, Brehmsstrahlung cooling and Compton cooling/heating.
For details of the rate coefficients and other atomic data that we use, see \citetalias{sykesFluorescentRingsStarfree2019} and \citet{cookePrimordialAbundanceDeuterium2016}.

We proceed iteratively, using the temperature profile to recompute the pressure profile and repeating the above procedure until a convergence criterion is met: namely, that the fractional ionizations in every radial cell change by less than 0.1\% between successive iterations.
With a converged ionization structure found, we then compute the volume emissivity $\varepsilon_\nu$ for an emission line with frequency $\nu$ as:
\begin{equation}
    \varepsilon_\nu(r) = h\nu\,n_{\text{ion}}(r)\,n_e(r)\,\alpha_{\text{eff}}(T(r)),\hfill%
    [\varepsilon]=\text{erg}\,\text{s}^{-1}\,\text{cm}^{-3}
    \label{eq:emis}
\end{equation}
where $h$ is Planck's constant, $n_{\text{ion}}$ is the density of $\HII$, $\HeII$ or $\HeIII$ as appropriate, $n_e$ is the electron density, and $\alpha_{\text{eff}}$ is a temperature-dependent coefficient expressing the rate per unit ion and electron densities at which the relevant transitions occur.
For $\Ha$ and the analogous $\HeII\;4686\aunit$ line, we obtain $\alpha_{\text{eff}}$ values from \citet{osterbrockAstrophysicsGaseousNebulae2006}, while for the $\HeI$ lines we use the emissivities compiled by
\citet{porterImprovedHeEmissivities2012,porterErratumImprovedHe2013}, which are tabulated as functions of $n_e$ and $T$.
Finally, we calculate the projected surface brightness $\Sigma_\nu$ as the integral of $\varepsilon_\nu(r)$ along lines of sight corresponding to an impact parameter $b$:
\begin{equation}
\Sigma_\nu(b)=\frac{1}{2\pi} \int_b^{\mathrlap{\Rv}}{\frac{r}{\sqrt{r^2 - b^2}}}\varepsilon_\nu(r)\,\df{r}\hfill%
[\Sigma]=\sbunit
\end{equation}

\section{Results}
\label{sec:results}

We first discuss the qualitative properties of the emission lines using a fiducial model, which takes the primordial helium abundance to be $\primHe=0.083$, corresponding to a primordial mass fraction $\primHeMass=0.249$.
\begin{figure}
    \centering
    \graphiccommand{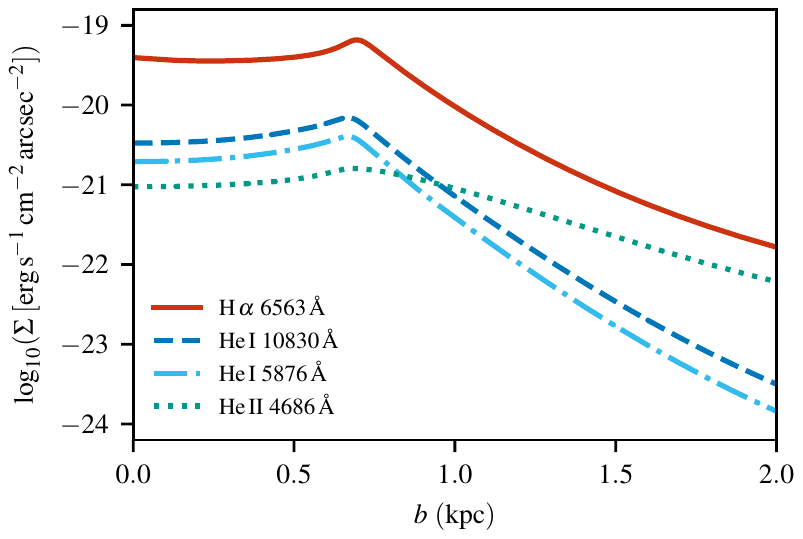}
    \caption{Surface brightness $\Sigma$ as a function of impact parameter $b$, for the $4686\aunit$ $\HeII$ line and two $\HeI$ lines, in addition to the hydrogen $\Ha$ line.}
    \label{fig:he_sb}
\end{figure}
In Fig.~\ref{fig:he_sb}, we show surface brightness profiles as a function of impact parameter for a halo with $\Mv=10^{9.55}\,\Msun$.
This is the maximum halo mass such that upper bounds on the column density and gas mass fraction, motivated by the need to avoid star formation, are not exceeded (see \citetalias[\S 2.3]{sykesFluorescentRingsStarfree2019}).
In addition to the hydrogen $\Ha$ line, we plot the surface brightness of the two brightest $\HeI$ lines (at $10830\aunit$ and $5876\aunit$), and the $\HeII\;4686\aunit$ line.
For this halo, the $\Ha$ surface brightness reaches a peak intensity of $\Sigma_{\text{\Ha, max}}=6.58\times10^{-20}\,\sbunit$.
The helium line surface brightnesses are significantly dimmer, reaching maximum values of $4.06\times 10^{-21}$ and $1.61\times 10^{-21}\,\sbunit$ for the $\HeI$ and $\HeII$ lines, respectively.

$\Sigma_{\text{\Ha, max}}$ occurs at $b=0.70\,\text{kpc}$, which corresponds to the projected radius of the fluorescent ring.
This location is set by the position of the hydrogen ionization front, at which the $\Ha$ volume emissivity reaches a maximum.
Outside the ionization front, $\varepsilon_{\Ha}$ falls rapidly with increasing radius as the gas density drops.
Conversely, at radii within the ionization front the emissivity is suppressed by the exponentially-decreasing ionized fraction.\footnote{At even smaller radii, $\varepsilon_{\Ha}$ begins to rise again due to an increased contribution from secondary collisional ionizations.}
We find that the peak helium surface brightnesses occur at a similar radial position to $\Sigma_{\text{\Ha, max}}$, despite the respective ionization fronts being located at different radii.
This occurs because the requirement of hydrostatic equilibrium produces gas densities which decrease rapidly with radius, such that the helium emissivities are affected more strongly by the falling electron density than by the helium ion densities.

From Eq.~\ref{eq:emis}, we would expect that the ratio of helium to hydrogen emissivity is set by the product of the ratios of ion densities, line frequencies and rate coefficients.
However, recovering the helium abundance $\primHe$ from this ratio involves some additional considerations.
Firstly, the densities in Eq.~\ref{eq:emis} refer to single ionization stages, whereas $\primHe$ is set by the overall atomic abundances.
Secondly, the rate coefficients $\alpha_{\text{eff}}$ are temperature-dependent.
Since the temperature and ionization structure of the gas (which influence $\alpha_{\text{eff}}$ and the fractional ionizations respectively) depend both on $\primHe$ and each other, we proceed by performing the iterative procedure outlined in Section~\ref{sec:model} for a number of different $\primHe$ values.
By comparing the results obtained in each case, we may determine the effects of changing $\primHe$ in relative terms. 

\subsection{Determining \texorpdfstring{$\primHe$}{y\_P}}
\label{ssec:yp}

To quantify how varying $\primHe$ changes the predicted emission (and hence the sensitivity for determining $\primHe$), we first consider the ratio of $\Ha$ to helium emissivity $\emr$, defined as:
\begin{equation}
  \emr\equiv\frac{\varepsilon(\Ha)}{\varepsilon(\HeII\,4686\aunit)+\varepsilon(\HeI\,10830\aunit)+\varepsilon(\HeI\,5876\aunit)},
  \label{eq:em_ratio}
\end{equation}
such that the net $\HeI$ emissivity is the sum of the brightest two $\HeI$ lines.
These comprise the near-infrared $10830\aunit$ line and the $5876\aunit$ line, which is intrinsically fainter but lies in the optical part of the spectrum along with the $\Ha$ and $\HeII$ lines, and so may be more convenient to detect. 
In addition to our fiducial model, we consider variations in which the assumed value of $\primHe$ is altered by a factor $f_y$:
\begin{equation}
    f_y\equiv \frac{\primHe}{\primHe_{,\,\text{fid}}},
    \label{eq:fy}
\end{equation}
which we allow to take the values $f_y=(1.01, 0.99, 1.10, 0.90)$, corresponding to $\pm 1\%$ and $\pm 10\%$ changes in $\primHe$.

Calculations for a fixed halo mass are not directly comparable between these variations, since the assumed helium abundance affects the thermal and ionization state of the gas, altering the conditions for hydrostatic equilibrium and leading to gas distributions with different emission properties.
We instead require that our results should be ``self-similar'' across model variations, in the sense that a change in the input parameter which distinguishes the variations should cause the predicted emission properties to change in a systematic way.
In \citetalias{sykesFluorescentRingsStarfree2019} we found that this condition is satisfied if we compare haloes whose gas content reaches the same peak column density of neutral hydrogen $N_{\HI,\,\text{max}}$.
This condition is appropriate because $N_{\HI,\,\text{max}}$ provides a proxy for $\tau_\infty$, the total optical depth of the gas.
This sets the intensity of the radiation field near the centre of the cloud, and thus its ionization state, in a way that is largely independent of the overall density and temperature structure, which does differ between model variations.
We employ the same approach here.
However, the relation between $\Mv$ and $N_{\HI,\,\text{max}}$ is not known\emph{ a priori}, so for each model variation we use an iterative procedure to determine the halo mass which results in the desired value of $N_{\HI,\text{max}}$. 
While we must choose a value of $N_{\HI,\text{max}}$ to enable comparisons between our model variations, the results we will present are insensitive to the threshold chosen, provided that it is sufficiently high that a well-defined ionization front is formed ($N_{\HI,\text{max}}>10^{19}\;\pcm$).
We adopt $N_{\HI,\text{max}}=10^{23}\,\pcm$ which, although large, is unlikely to result in self-shielded gas becoming star-forming due to the extremely metal-poor nature of the gas, which results in inefficient cooling.
Furthermore, the steep decline of $N_{\HI}$ with radius means that this column density is achieved only for sightlines passing through the very centre of the RELHIC.
\footnote{See Section~\ref{sec:discuss} of this paper and Section 2.3 of \citetalias{sykesFluorescentRingsStarfree2019} for further discussion.}

\begin{figure}
    \centering
    \graphiccommand{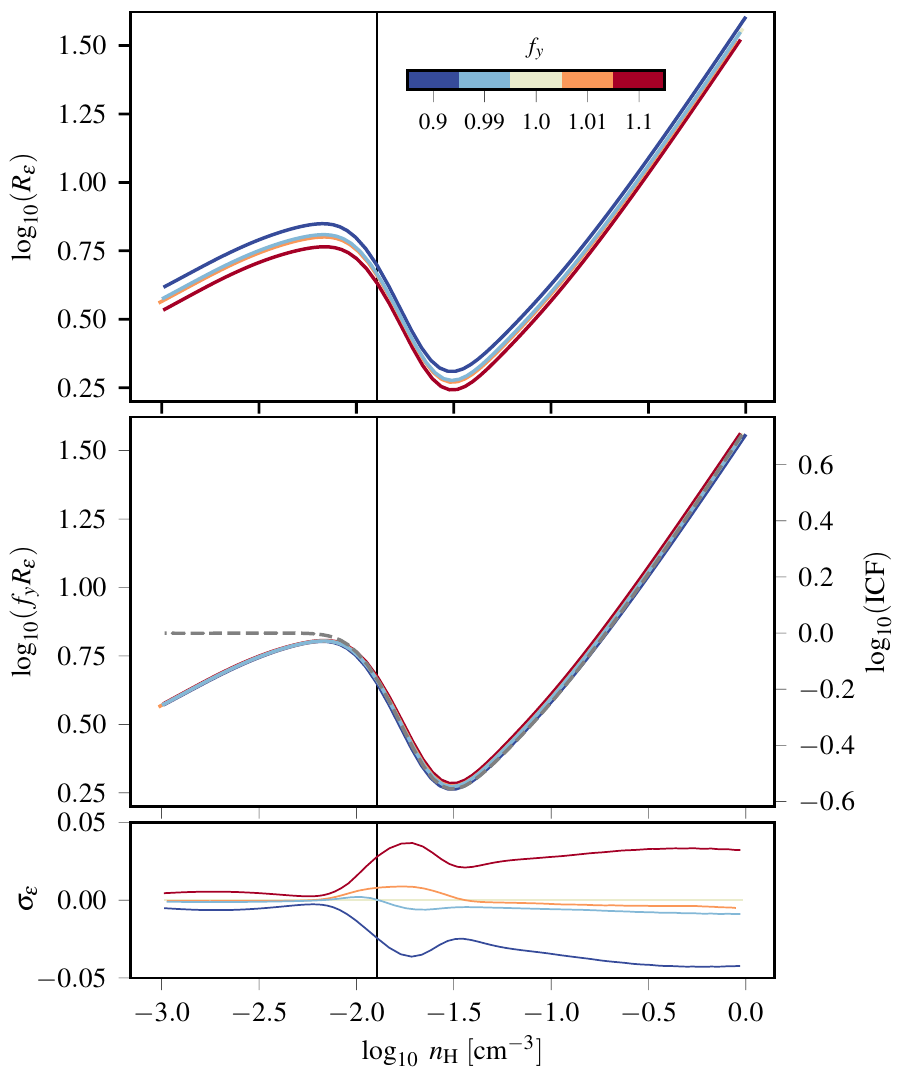}
    \caption{\emph{Upper panel}: Hydrogen to helium emissivity ratio $\emr$ (Eq.~\ref{eq:em_ratio}) for models with different $f_y$ (Eq.~\ref{eq:fy}), as a function of $n_{\text{H}}$. \emph{Middle panel}: Emissivity ratios, scaled additionally by $f_y$, for the same models. The ionization correction factor for $f_y=1$, defined in the text, is shown with a dashed grey line using the right-hand scale. \emph{Lower panel}: Deviation from self similarity $\sigma_\varepsilon$ (Eq.~\ref{eq:em_sigma}), i.e. the error incurred in assuming $f_y \emr$ is independent of $\primHe$. The solid vertical lines indicate the density at the location of the ionization front.}
    \label{fig:icf_emis}
\end{figure}
We plot the resulting emissivity ratios, as a function of the hydrogen number density $n_{\text{H}}$, in the upper panel of Fig.~\ref{fig:icf_emis}.
As expected, $\emr$ decreases (i.e. $\varepsilon_{\text{He}}$ is larger relative to $\varepsilon_{\Ha}$) for models with a larger value of $\primHe$, and conversely for smaller $\primHe$.
In the middle panel of Fig.~\ref{fig:icf_emis}, we additionally scale the obtained $\emr$ curves by $f_y$ to demonstrate that these profiles remain almost self-similar, particularly near the ionization front where the emissivities are maximised.
We quantify this property in the lower panel, where the fractional deviation from self-similarity $\sigma_\varepsilon$, defined as:
\begin{equation}
\sigma_\varepsilon\equiv\frac{f_y \emr - \emr_{\text{\!,\,fid}}}{\emr_{\text{\!,\,fid}}},
\label{eq:em_sigma}
\end{equation}
is plotted for each value of $f_y$.
We find larger values of $\sigma_\varepsilon$ at higher densities, corresponding to gas located inside the ionization front.
Identifying the causes of these deviations is made difficult by the coupled nature of our calculations, but we expect differences in the thermal state and density profile of the gas (in particular, the electron density $n_{\text{e}}$) between models to be important.
Experiments where the variation in $\primHe$ was ignored when determining these quantities reduced, but did not eliminate, the residual discrepancies.
In any case, they represent deviations at the position of the peak emissivity of ${<}1\%$ for $1\%$ changes in $\primHe$ and ${\sim}2\%$ for $10\%$ changes.
Hence, they may be safely discounted, and we are justified in interpreting the variations in $\emr$ as being solely caused by changes in $\primHe$.
This direct relationship between the observable values of $\emr$ and the underlying abundance is the fundamental property that makes RELHICs appealing tools for determining $\primHe$.

The emissivities and their ratio depend on the densities of a specific ionization stage, whereas determining $\primHe$ requires the total densities of H and He.
This scenario is commonly encountered in absorption and emission line studies, and is circumvented by introducing an ionization correction factor (ICF) to account for unobserved ionization stages, allowing the total density to be inferred.
We define the ICF as follows:
\begin{equation}
    \ICF \equiv \frac{n_{\HII}}{n_{\text{H}}} \frac{n_{\text{He}}}{n_{\HeII} + n_{\HeIII}} = \frac{1-X_{\HI}}{1-X_{\HeI}},
    \label{eq:icf}
\end{equation}
noting that this expression has a similar functional form to the emissivity ratio (by using Eq.~\ref{eq:emis} to expand each of the terms in Eq.~\ref{eq:em_ratio}).
We illustrate this correspondence in the middle panel of Fig.~\ref{fig:icf_emis}, where the ICF is shown by the grey dashed line and plotted on the right-hand axis.
The equilibrium temperature of the gas falls with decreasing density, which in combination with the different temperature dependencies of the recombination coefficients, cause $\emr$ to also fall with decreasing density.
Conversely, at low densities both H and He become fully ionized, leading to the ICF tending to 1 and breaking the correspondence with $\emr$.
Nevertheless, at densities of $n_H\sim 10^{-2}\,\text{cm}^{-3}$ associated with the peak emissivity, assuming a direct proportionality between $\emr$ and the ICF allows a measurement of the former to be translated to an ionization correction, and hence a value of $\primHe$, to a precision of within 5\%.

However, the volume emissivity is not itself an observable quantity, but rather its integral along the line of sight, the surface brightness.
For situations in which the gas distribution can be described by a plane-parallel model, this distinction is not problematic.
In the spherical geometry that we consider here, projection effects are significant since the surface brightness peak occurs due to limb brightening along lines of sight passing through more strongly-emitting gas.
To determine the impact of these projection effects, we define the surface brightness ratio, analogously to $\emr$, as:
\begin{equation}
  \sbr\equiv\frac{\Sigma(\Ha)}{\Sigma(\HeII\,4686\aunit)+\Sigma(\HeI\,10830\aunit)+\Sigma(\HeI\,5876\aunit)}.
  \label{eq:sb_ratio}  
\end{equation}
We also define the deviation from surface brightness self-similarity as:
\begin{equation}
\sigma_\Sigma\equiv\frac{f_y \sbr - \sbr_{\text{,\,fid}}}{\sbr_{\text{,\,fid}}}.
\label{eq:sb_sigma}
\end{equation}
\begin{figure}
    \centering
    \graphiccommand{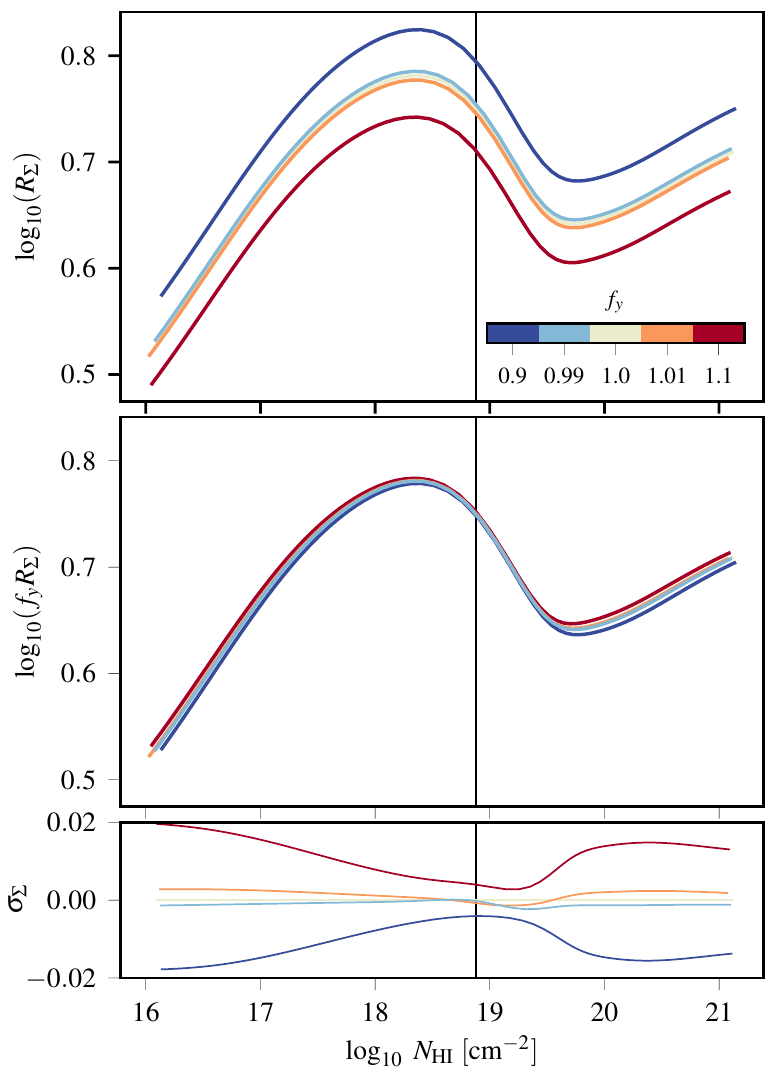}
    \caption{\emph{Upper panel:} Hydrogen to helium surface brightness ratio $\sbr$ (Eq.~\ref{eq:sb_sigma}) for the same models shown in Fig.~\ref{fig:icf_emis}, as a function of $N_{\text{\HI}}$. \emph{Middle panel:} Surface brightness ratios scaled by $f_y$. \emph{Lower panel}: Deviation from self similarity $\sigma_\Sigma$ (Eq.~\ref{eq:sb_sigma}), i.e. the error incurred in assuming $f_y\sbr$ is independent of $\primHe$. The solid vertical lines indicate the $\HI$ column density at the location of the ionization front.}
    \label{fig:icf_sb_ratio}
\end{figure}
We plot $\sbr$, $f_y \sbr$, and $\sigma_\Sigma$ for the same models shown previously in Fig.~\ref{fig:icf_sb_ratio}, where the $x$-axis now shows projected neutral hydrogen column densities $N_{\HI}$ (rather than $n_{\text{H}}$ as used in Fig.~\ref{fig:icf_emis}).
We find that despite the projection effects, self-similarity is closely preserved when moving to surface brightnesses, with values of $\sigma_\Sigma$ at the position of the peak surface brightness in fact being smaller than the equivalent quantity for $\emr$. 
This is likely due to the fact that the surface brightness at any radius is calculated by integrating over the entire emissivity profile, allowing for a degree of fortuitous cancellation between errors in opposite directions.
However, determining $\primHe$ from $\sbr$ requires accurate measurements of the peak surface brightness $\Sigma_{\text{max}}$ in $\Ha$ and the three helium lines we consider.
In absolute terms this emission is still extremely faint, particularly for the $\HeII\;4686\aunit$ line for which $\Sigma_{\text{max}}\sim 10^{-21}\,\sbunit$.
Consequently, obtaining a measurement of the surface brightness with the precision needed to produce a competitive measurement of $\primHe$ would be very challenging using current instrumentation.
The integrated nature of the surface brightness may also impact our results, since it means that our predicted values of $\sbr$ are sensitive to our modelling of the complete temperature and density structure of the gas, whereas $\emr$ depends only on local values of $T$ and $n_e$.
However, we do not expect this to be a significant disadvantage, given that temperature and density structures for RELHICs are well-specified.

We now consider the possibility of inferring $\primHe$ from the total line flux across the projected area of the fluorescent ring.
This is observationally more feasible since measuring fluxes does not depend on making a highly-precise measurement of the peak surface brightness.
Hence the fluorescent ring itself need not be spatially resolved, and the measurement precision attainable depends solely on the precision with which the flux can be determined (the signal-to-noise ratio (SNR) of the observations).
Moreover, the flux will depend only on the well-specified total gas content of the halo, given that the bulk of the emission originates in optically-thick gas for which the intensity of emission may be predicted analytically \citep[see also App. A, \citetalias{sykesFluorescentRingsStarfree2019}]{gouldImagingForestLyman1996}.
We integrate the surface brightness over impact parameter to calculate total line luminosities, finding typical values of $L_{i}\approx (6\times10^{35},\,9\times10^{34},\,3\times10^{34})\,\text{erg}\,\text{s}^{-1}$ for the $\Ha$, total $\HeI$, and $\HeII\,4686\aunit$ luminosities respectively.
Converting these luminosities to line fluxes would require assuming a distance from the observer to the RELHIC that we model.
However, we wish to consider the ratio of hydrogen and helium fluxes, which remains distance-independent and will be equal to the ratio of the total line luminosities.
Hence, we define the $\Ha$ to helium flux ratio as:
\begin{equation}
  \lr\equiv\frac{F(\Ha)}{F(\HeII\,4686\aunit)+F(\HeI\,10830\aunit)+F(\HeI\,5876\aunit)},
  \label{eq:lum_ratio}
\end{equation}
but in practice, compute the luminosity ratio $R_{\text{L}}$ instead.
\begin{figure}
    \centering
    \graphiccommand{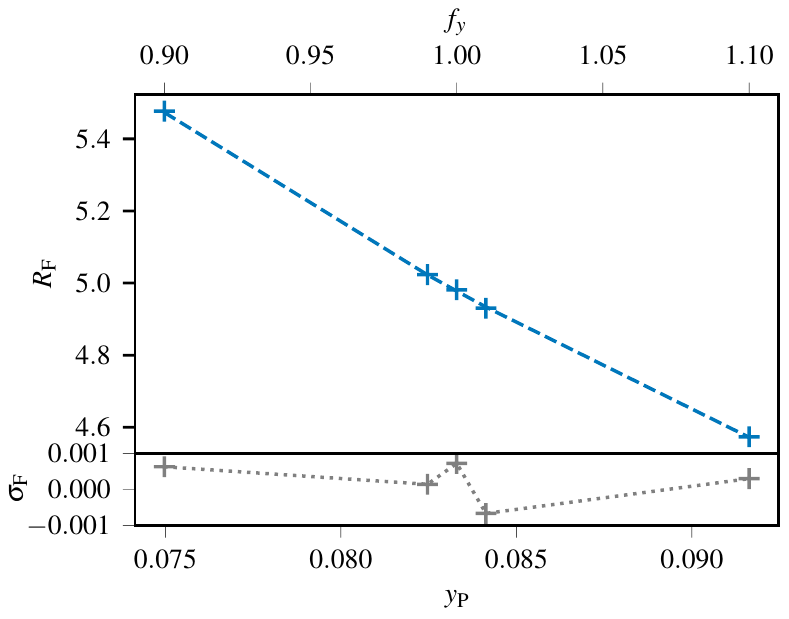}
    \caption{\emph{Upper panel}: Flux ratio $\lr$ (Eq.~\ref{eq:lum_ratio}) as a function of $\primHe$.
    The dashed line indicates a fit of the form $\lr=A/\primHe+b$, where $A=0.372$ and $b=0.516$ are arbitrary scaling constants.
    \emph{Lower panel}: The relative error $\sigma_F$ (Eq.~\ref{eq:lum_sigma}) between the calculated flux ratios and the fitted line.}
    \label{fig:lum_ratio}
\end{figure}
In the main panel of Fig.~\ref{fig:lum_ratio} we show the values of $\lr$ we obtain for different values of $y_p$, finding the expected trend of decreasing $\lr$ with increasing $\primHe$.
More quantitatively, we expect that $\lr\propto 1/\primHe$, with the normalisation of this relation being set by the relative intrinsic emissivities of the three lines we consider.
We use standard non-linear least squares regression to fit a curve of this functional form to the predicted values of $\lr$, shown by the dashed curve in Fig.~\ref{fig:lum_ratio}.
To improve the fit, we additionally compute values of $\lr$ for $f_y=\pm 20\%$ (not shown in Fig.~\ref{fig:lum_ratio}).
We define $\sigma_{\text{F}}$ as the normalised residual of the data with respect to this fit:
\begin{equation}
    \sigma_F\equiv\frac{\lr - \lr_{\text{\!,\,fit}}}{\lr_{\text{\!,\,fit}}},
    \label{eq:lum_sigma}
\end{equation}
and plot this as a function of $\primHe$ in the lower panel, finding that $\sigma_{\text{F}}\ll1\%$ over the range of $\primHe$ values we consider.

We next use this fit to determine the precision with which $\primHe$ may be inferred from measuring $\lr$, given that measurements of the fluxes from which the flux ratio is calculated will have an associated uncertainty.
We assume that this uncertainty is described by a single relative error value $\lerr$ for each of the four fluxes that must be measured, and use standard error propagation to determine the resulting error in $\lr$.
By inversion of the fit in Fig.~\ref{fig:lum_ratio}, we obtain a range of values of $\primHe$ consistent with the imprecise value for $\lr$, the extrema of which we report as $\yperr$, the error in $\primHe$.
\begin{figure}
    \centering
    \graphiccommand{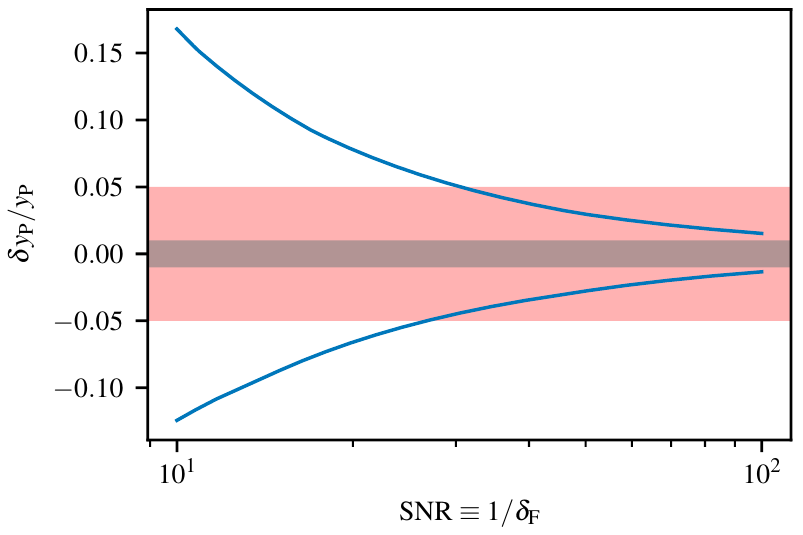}
    \caption{The relative precision to which $\primHe$ can be inferred (blue curves) as a function of $\lerr$, the precision to which the individual line fluxes are measured. The true helium abundance is assumed to be $\primHe=0.083$.
    Red and grey shaded bands show 5\% and 1\% errors on $\primHe$ respectively.}
    \label{fig:lum_ratio_snr}
\end{figure}
In Fig.~\ref{fig:lum_ratio_snr}, we plot normalised values of $\yperr$ as a function of the flux SNR, defined as $\text{SNR}=1/\lerr$.
We find that for SNRs of 10 and 100, corresponding to $\lerr=10\%$ and $1\%$, $\primHe$ may be inferred to a precision of ${}^{+13}_{-10}\%$ and ${}^{+1.2}_{-1.0}\%$ respectively, where we have assumed that the underlying `true' helium abundance is the fiducial value $\primHe=0.083$; repeating these calculations assuming different values of $\primHe$ does not significantly affect the obtained values of $\yperr$.
The asymmetry in these limits results from the non-linearity of the function $\lr(\primHe)$, which means that for a flux error of fixed magnitude $|\lerr|$, the magnitude of $\yperr$ will vary depending on the sign of $\lerr$.
Fig.~\ref{fig:lum_ratio_snr} may also be used to determine the SNR required to achieve a constraint on $\primHe$ of a given precision.
The 5\% constraint indicated by the red shaded region corresponds to the range of reported values of $\primHe$ (see references in Section~\ref{sec:intro}); in the absence of a uniform systematic offset in these measurements, this is the minimum level of precision which must be reached for an independent measurement to provide additional information.
Conversely, the grey region indicates a 1\% constraint, as obtained by the most precise determinations of $\primHe$ currently available~\citep[e.g.][]{valerdiDeterminationPrimordialHelium2019}.
We find that satisfying these two constraints requires a flux SNR of ${\sim}30$ and ${\sim}140$ respectively.
Hence, measurements of $\lr$ from RELHICs have the potential to provide competitive constraints on $\primHe$, provided that the individual emission line fluxes can be determined to a precision of $\lerr\approx3\%$ or better.
As discussed, we expect this conclusion to be insensitive to the details of the gas distribution given that it remains optically thick.

\subsection{Determining the UVB spectral slope}
\label{ssec:aUV}

In Section~\ref{ssec:yp}, we assumed that the UVB is known (and is given by the \citetalias{madauCosmicReionizationPlanck2015} spectrum) in order to identify the effects of varying $\primHe$ in isolation.
In reality, the UVB spectral shape is poorly constrained at $z\sim 0$, as demonstrated by the variance between different UVB synthesis models~\citep[see e.g.][]{puchweinConsistentModellingMetagalactic2019,madauCosmicReionizationPlanck2015,faucher-giguereNewCalculationIonizing2009}.
These discrepancies may be further compounded by the uncertain impact of inhomogeneities in the UVB resulting from local sources.
In \citetalias{sykesFluorescentRingsStarfree2019}, we explored the effects of varying the UVB spectral slope on the properties of $\Ha$ rings, finding that a harder UVB produced brighter rings at higher characteristic halo masses, and vice versa for a softer UVB.
The properties of helium rings will also be affected by the UVB slope, since a harder spectrum contains a greater proportion of helium-ionizing photons and will therefore produce brighter helium emission at fixed $\primHe$.
As discussed in Section~\ref{sec:intro}, existing measurements of $\primHe$ approach a precision of 1\%, and so it is reasonable to take this value as exact, and instead use the observable properties of helium rings to infer the UVB slope.

As in \citetalias{sykesFluorescentRingsStarfree2019}, we parameterise the UVB slope using the shape parameter $\alpha_{\rm{UV}}$ introduced by \citet{crightonMetalenrichedSubkiloparsecGas2015}, which modifies the slope of a given reference spectrum as follows:
\begin{equation}
J_{\nu}(E)=\begin{cases}
N_{\Gamma} \times J_{\nu,\text{ref}}(E) & E \leq E_0\\
N_{\Gamma} \times J_{\nu,\text{ref}}(E) \times (E/E_0)^{\alpha_{\rm{UV}}} & E_0 < E \leq E_1\\
N_{\Gamma} \times J_{\nu,\text{ref}}(E) \times (E_1/E_0)^{\alpha_{\rm{UV}}} & E > E_1,
\end{cases}
\end{equation}
where $J_{\nu,\text{ref}}(E)$ is the mean intensity of the reference spectrum at energy $E$, and $E_0$ and $E_1$ are pivot points between which we modulate the mean intensity by an additional power law with exponent $\aUV$.
We set $E_0=1\,\text{Ryd}$ and $E_1=10\,\text{Ryd}$, and introduce an additional factor $N_{\Gamma}\equiv\Gamma_{\HI,\text{fid}}/\Gamma_{\HI}$.
This acts to renormalise the spectra such that the $\HI$ photoionization rate of the modified UVB is the same as that of the reference spectrum, which for our purposes is the \citetalias{madauCosmicReionizationPlanck2015} UVB.
In addition to our fiducial model which corresponds to $\aUV=0$, we compute models for RELHICs illuminated by UVBs with $\aUV=(-2, -1, 1)$.
These values are chosen to cover all realistic UVB spectra between the extremes of a soft, starburst-driven spectrum ($\aUV=-2$) and a hard, AGN-dominated spectrum ($\aUV=1$).
As described at the start of Section~\ref{sec:results}, for each value of $\aUV$ we iteratively perform calculations for haloes of different masses to obtain a model RELHIC with $\HI$ column density equal to the adopted threshold $N_{\HI,\text{max}}=10^{23}\;\pcm$.
\begin{figure}
    \centering
    \graphiccommand{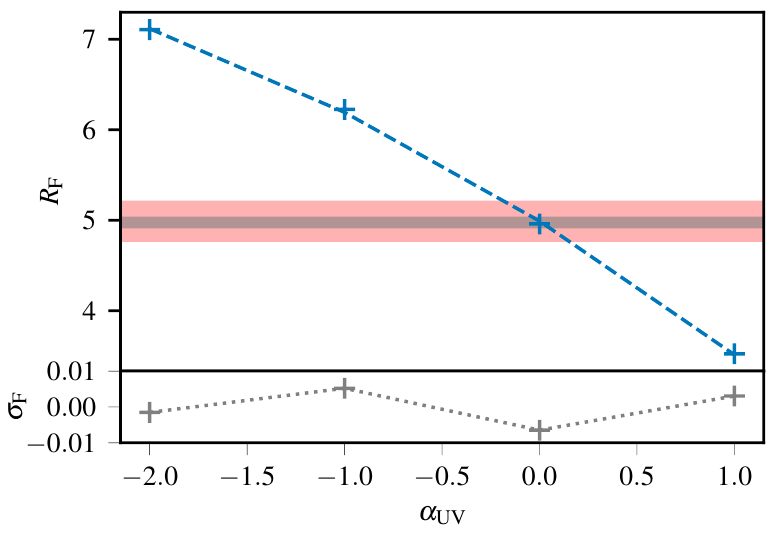}
    \caption{\emph{Upper panel}: Flux ratio $\lr$ as Fig.~\ref{fig:lum_ratio}, but for models with  fiducial value $\primHe=0.833$ and different UVB slopes $\aUV$.
    The dashed line indicates a quadratic fit $\lr=A\aUV^2+B\aUV+C$, where $A=-0.138$, $B=-1.34$ and $C=4.99$.
    Red and grey shading indicates the range of $\lr$ values consistent with $\pm 5\%$ and $\pm 1\%$ variations of $\primHe$ respectively.
    \emph{Lower panel}: The relative error $\sigma_F$ between the calculated flux ratios and the fitted line.}
    \label{fig:lum_ratio_aUV}
\end{figure}
We calculate values for $\lr$ as described previously, and plot these as a function of $\aUV$ in Fig.~\ref{fig:lum_ratio_aUV}.
In red (grey) shading, we show the range of values of $\lr$ resulting from ${\pm} 5\%$ ($1\%$) variations in $\primHe$, as plotted in Fig.~\ref{fig:lum_ratio_snr}.
We see that changing $\aUV$ results in a much wider range of $\lr$ values than changing $\primHe$.
Thus, if the UVB slope deviates significantly from that of the \citetalias{madauCosmicReionizationPlanck2015} spectrum, $\lr$ will change from its fiducial value by a greater margin than could be caused by any reasonable uncertainty in $\primHe$, allowing the two effects to be distinguished.

We use a quadratic fit to describe the variation of $\lr$ with $\aUV$, which reproduces the data to a precision of $1\%$ or better, as shown in the bottom panel of Fig.~\ref{fig:lum_ratio_aUV}.
In the same manner as was done for $\primHe$, we use this fit to compute the precision with which $\aUV$ may be reconstructed from uncertain measurements of $\lr$.
\begin{figure}
    \centering
    \graphiccommand{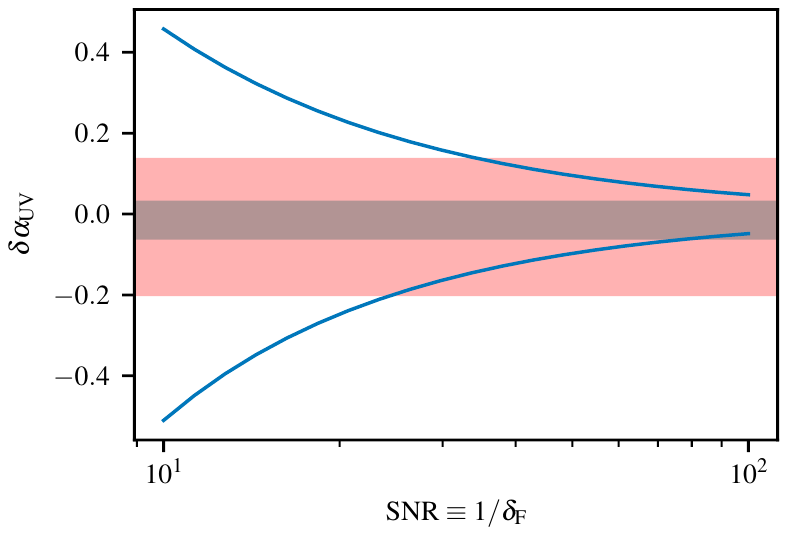}
    \caption{The absolute error with which $\aUV$ can be inferred as a function of $\lerr$. The true UVB slope parameter is assumed to be $\aUV=0$. Red and grey shading indicates the values of $\aerr$ for which the expected change in $\lr$ is degenerate with that caused by changes in $\primHe$ of 5\% and 1\% respectively.}
    \label{fig:lum_ratio_aUV_snr}
\end{figure}
This is shown in Fig.~\ref{fig:lum_ratio_aUV_snr}, where we plot $\aerr$, the absolute measurement uncertainty in $\aUV$, as a function of $\lerr$.
We find that SNRs of 10 and 100 yield constraints on $\aUV$ of ${}^{+0.46}_{-0.51}$ and ${}^{+0.048}_{-0.048}$ respectively.
The red and grey shaded regions now show the range of values of $\aerr$ for which the corresponding values of $\lr$ could also be obtained in models with $\aUV=0$ and $f_y \neq 1$. 
Thus, if the underlying UVB slope deviates from the fiducial value by an amount $\aerr\lesssim 0.2$, the expected change in $\lr$ is degenerate with that attributable to ${\pm}5\%$ changes in $\primHe$.

\subsection{Combined constraints on \texorpdfstring{$\primHe$}{y\_p} and \texorpdfstring{$\aUV$}{α\_UV}}
\label{ssec:combined}

We have demonstrated that it is feasible to determine either $\primHe$ or $\aUV$ using the hydrogen-to-helium flux ratio $\lr$, assuming perfect knowledge of the other property.
However, this is not representative of the real-world scenario in which both $\primHe$ and $\aUV$ are uncertain, as illustrated by the degeneracy visible in Fig.~\ref{fig:lum_ratio_aUV_snr} and discussed above.
In this section we investigate the possibility of simultaneously constraining $\primHe$ and $\aUV$.

\begin{figure}
    \centering
    \graphiccommand{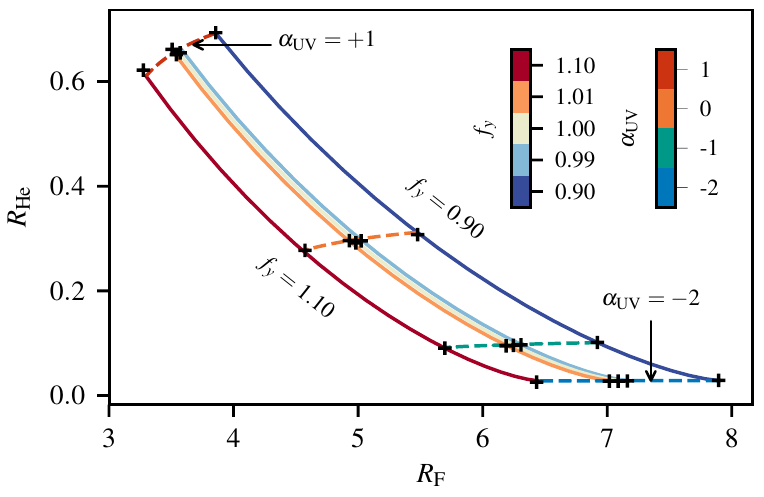}
    \caption{$\lr$ (Eq.~\ref{eq:lum_ratio}) vs. $\her$ (Eq.~\ref{eq:he_ratio}) for models with different $\primHe$ and $\aUV$. Black points indicate values calculated from runs of our ionization balance code; curves show best fits to this data using Eqs.~\ref{eq:2d_fits1} and \ref{eq:2d_fits2}.
    Curves at constant $\primHe$ and different values of $\aUV$ are shown with solid lines, coloured according to the left-hand colourbar.
    Conversely, curves at constant $\aUV$ and varying $\primHe$ are shown as dashed lines and coloured according to the right-hand colourbar.}
    \label{fig:lum_ratio_2d_scatter}
\end{figure}

In order to do this, it is necessary to break the degeneracy between $\primHe$ and $\aUV$, which both influence the value of $\lr$.
Since harder UVB spectra will contain more $\HeII$-ionizing photons, we expect the ratio
\begin{equation}
    \her\equiv\frac{F(\HeII\,4686\aunit)}{F(\HeI\,10830\aunit) + F(\HeI\,5876\aunit)}
    \label{eq:he_ratio}
\end{equation}
to increase with increasing $\aUV$.
In contrast, changing $\primHe$ scales all the helium ionic abundances equally, and so will only have a minor, indirect effect on $\her$ arising from the slight change to the abundance of free electrons, which affects the $\HeI$ and $\HeII$ emissivities differently.
We supplement our existing models, which vary either $\primHe$ or $\aUV$ while keeping the other parameter constant, with additional runs of our photoionization code in which both $\primHe$ and $\aUV$ are varied.
We calculate values of $\lr$ and $\her$ for this grid of models, which we show as the black points in Fig.~\ref{fig:lum_ratio_2d_scatter}.
As in the one-dimensional cases presented previously, we next fit a 2D surface to the calculated flux ratios, in order to allow interpolation of the ratios for arbitrary values of $\primHe$ and $\aUV$. We define these fits as follows:
\begin{align}
    \lr&=f_1(\primHe)\;g_1(\aUV)\label{eq:2d_fits1}\\
    \her&=f_2(\primHe)\;g_2(\aUV)\label{eq:2d_fits2},
\end{align}
where $g_1$, $g_2$ and $f_2$ are cubic polynomials, and $f_1\propto 1/\primHe$ as in Section~\ref{ssec:yp}.
We emphasise that these fits are intended to be empirical only, and are chosen for their simplicity.
Nevertheless, they are able to reproduce the flux ratios obtained from our simulations to an accuracy of 3\% or better across the range of $\primHe$ and $\aUV$ values we consider.
We evaluate these fits and plot the resulting curves in Fig.~\ref{fig:lum_ratio_2d_scatter} in order to show the degree to which they reproduce the data.
\begin{figure*}
    \centering
    \resizebox{\textwidth}{!}{\graphiccommand{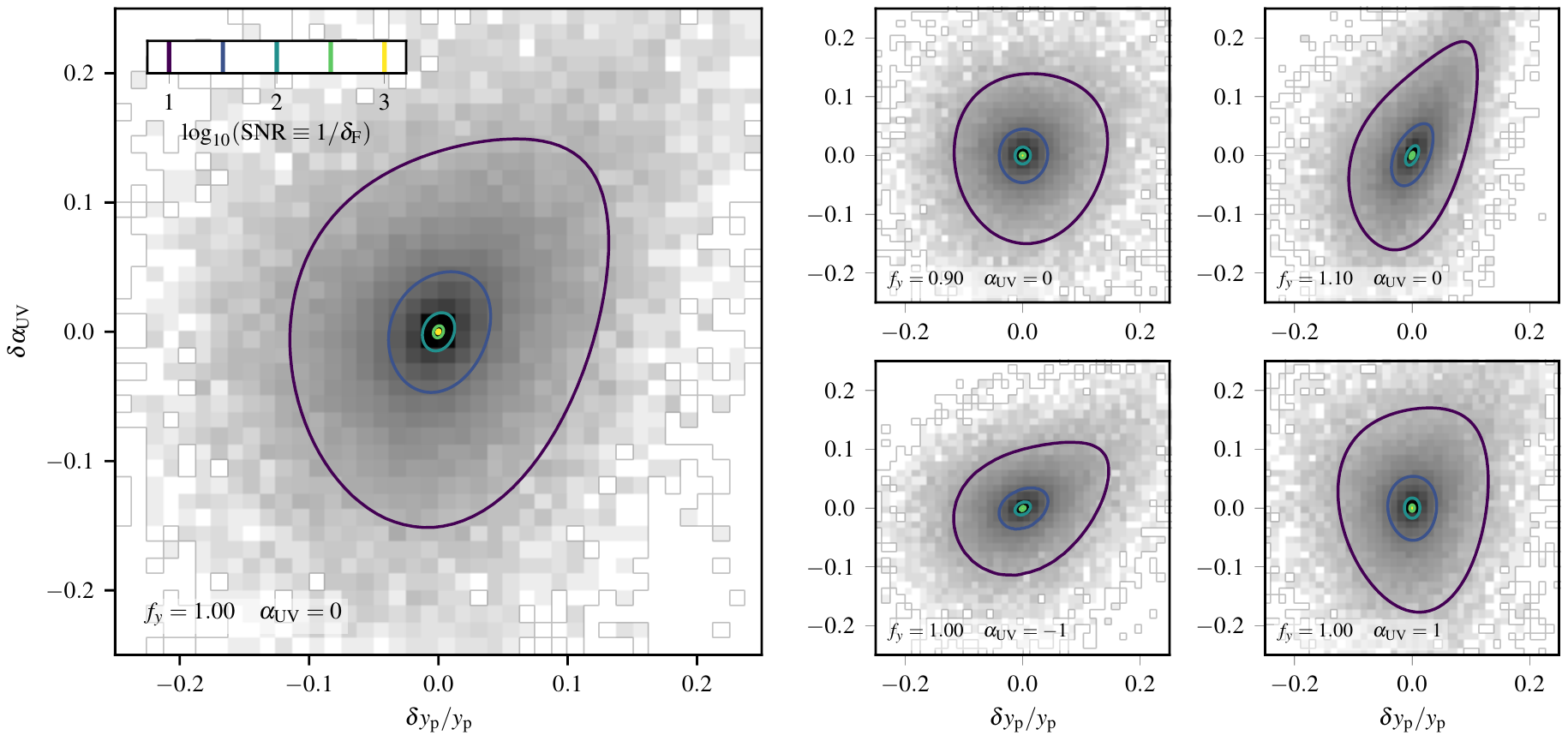}}
    \caption{\emph{Left panel}: Histogram of precision with which $\primHe$ and $\aUV$ may be recovered from sampled $\lr$ and $\her$ values.
    Bins are shaded according to their occupancy using a logarithmic scale.
    Contours indicate the precision obtained for $1\sigma$ flux measurement uncertainties in the range $0.001\leq\lerr\leq0.1$.
    The $y$-axis shows the absolute measurement uncertainty in $\aUV$, whereas the relative error is plotted for $\primHe$.
    \emph{Right panels}: As left, but for different underlying $\primHe$ and $\aUV$ values, as indicated by the legend in the bottom-left of each panel.} 
    \label{fig:lum_ratio_aUV_Yp_snr}
\end{figure*}

As a result of the non-linear mapping between the $\lr$--$\her$ and $\primHe$--$\aUV$ axes, we employ a Monte Carlo technique to estimate the precision with which the latter parameters may be recovered.
We begin by choosing the magnitudes of the uncertainties with which $\lr$ and $\her$ are measured, which we derive by choosing a single flux error $\lerr$ and propagating this uncertainty into $\lr$ and $\her$ as described previously.
Taking the underlying values of $\primHe$ and $\aUV$ to be 0.083 and 0 respectively, we sample 1000 uncertain `measurements' of $\lr$ and $\her$ by drawing from a bivariate normal distribution with means given by evaluating Eqs.~\ref{eq:2d_fits1} and \ref{eq:2d_fits2} for the true values.
Variances are set to the squares of the chosen flux ratio errors, such that these errors correspond to $1\sigma$ uncertainties.
We then invert Eqs.~\ref{eq:2d_fits1} and \ref{eq:2d_fits2} to map each of the sampled $\lr$--$\her$ values to the $\primHe$--$\aUV$ plane, and repeat this procedure for different values of $\lerr$.
In the left-hand panel of Fig.~\ref{fig:lum_ratio_aUV_Yp_snr}, we illustrate the resulting collection of $\yperr$--$\aerr$ samples using a two-dimensional histogram.
Also shown is a series of contours indicating the $1\sigma$ limits on $\primHe$ and $\aUV$ which result from different choices of $\lerr$ in the range $0.001\leq\lerr\leq0.1$.

We see that the joint constraints also provide comparable or better precision than the individual ones, particularly for $\aUV$. The addition of the $\her$ measurement permits an improvement in the reconstructed precision of this parameter by approximately a factor of 4, with $\lerr=0.1$ now yielding the constraint on $\aUV$ of ${\pm}0.15$.
Conversely, the precision with which $\primHe$ may be recovered is ${}^{+14}_{-12}\%$, effectively unchanged to that obtained from the individual constraints (Fig.~\ref{fig:lum_ratio_snr}).
These results are in agreement with the indication in Fig.~\ref{fig:lum_ratio_2d_scatter} that $\her$ evolves much more strongly with $\aUV$ than with $\primHe$.
Increasing the assumed measurement precision to $\text{SNR}=100$ improves these constraints significantly, to give ${\pm}1.3\%$ and ${\pm}0.015$ constraints on the helium abundance and UVB slope respectively.

In the right-hand panels, we repeat the Monte Carlo process outlined above, but using different underlying values of $\primHe$ and $\aUV$.
We obtain comparably precise constraints in all cases shown, with $\text{SNR}=100$ yielding values of $\yperr/\primHe$ of ${\pm}1.3\%$ or better, and $\aerr$ of $\pm0.017$ or better.
Furthermore, a SNR of $10^{1.5}\approx32$, corresponding to the second-outermost contour, is always sufficient to recover $\primHe$ to the 5\% level at which existing determinations of the abundance differ.
The tendency for a positive correlation between $\yperr$ and $\aerr$ is again a consequence of the general shape of Fig.~\ref{fig:lum_ratio_2d_scatter}: a positive value for $\yperr$ is produced by a negative absolute error on $\lr$.
When combined with an error on $\her$ which is also ${\geq}0$, the reconstructed value of $\aUV$ exceeds the assumed underlying value, and therefore $\aerr$ is positive also.
Equivalently, a positive error on $\lr$ and a negative error on $\her$ combine to yield inferred values of $\primHe$ and $\aUV$ that lie below the true ones.
Curves at constant $\aUV$ (defined parametrically by Eqs.~\ref{eq:2d_fits1} and \ref{eq:2d_fits2}) flatten toward lower $\primHe$, meaning this correlation is not as pronounced and giving the constraint contours shown in Fig.~\ref{fig:lum_ratio_aUV_Yp_snr} their ovoid shape.

\section{Summary and Conclusions}
\label{sec:discuss}



We have examined the properties of hydrogen and helium emission driven by UV background fluorescence in RELHICs, a class of ${\sim}10^{9.5}\,\Msun$ dark matter haloes which fail to form stars, instead retaining a small reservoir of neutral, essentially-pristine gas at redshift $z=0$.
Using results obtained from a specialised radiative transfer code, we have shown that the ratio of hydrogen to helium emission relates directly to the helium abundance of the gas.
In particular, we showed that from ratios of integrated quantities, such as the surface brightness and integrated flux, we are able to recover the assumed helium abundance to 1\% or better.
Hence, these measurements have the potential to provide an independent measurement of the primordial helium abundance.

RELHICs are intrinsically simple systems, making them ideally suited for assessing the presence of systematic errors in the canonical method for measuring $\primHe$ using metal-poor $\HII$ regions.
For example, the expected almost-pristine nature of the gas in RELHICs avoids the need to extrapolate observed helium abundance measures down to zero metallicity.
Their well-specified structure, in which the majority of emission is produced by gas which settles at its photoionization equilibrium temperature, and at a density dictated by the requirement of hydrostatic equilibrium, reduces the impact of systematics which can arise from uncertainties in the temperature and density structure of $\HII$ regions, as well as in the degree to which they are chemically homogeneous~\citep{izotovPrimordialAbundanceOf4He2007}.
Additionally, the much lower typical density of the gas within RELHICs means that all emissivities may be calculated fully in the low-density limit, which can result in up to an order-of-magnitude reduction in their associated uncertainties~\citep{porterUncertaintiesTheoreticalHe2009}.
Finally, this approach provides the added bonus of permitting the spectral slope of the $z=0$ UVB to be inferred, as discussed in Sections~\ref{ssec:aUV} and \ref{ssec:combined}.
Intrinsically `dark' sources like RELHICs are uniquely positioned to allow such a measurement to be made, as any attempt to infer the slope of the UVB using the nebular emission from luminous sources requires the subtle effect of the UVB to be disentangled from the effect of the locally-produced radiation field.

While the strengths of this approach are promising, significant challenges also exist.
We make a number of modelling assumptions, such as assuming RELHICs to be spherically-symmetric and in hydrostatic equilibrium with a gravitational potential due solely to their dark matter content.
\citetalias{benitez-llambayPropertiesDarkLCDM2017} reports that the first two assumptions are in agreement with the properties of RELHICs identified in \textsc{Apostle}, while neglecting the gas self-gravity is justified since $\Mv\gg M_{\text{g}}$ for these systems.
More significantly, while the existence of dark matter haloes in the mass range corresponding to RELHICs is a robust result of CDM structure formation, the prediction that they remain star-free but gas-rich is less certain.
The limited spatial and mass resolution of cosmological simulations means that they are unable to follow the formation of individual stars. 
In addition, following the physical processes governing the formation of a cold, molecular gas phase is computationally intensive.
Thus, the \textsc{Apostle} simulations instead enforce an effective equation of state for cool gas, and consider this gas to be eligible for star formation when it exceeds a metallicity-dependent density threshold, as proposed by \citet{schayeStarFormationThresholds2004}.
For the extremely low-metallicity gas RELHICs contain, this threshold is set to $n_{\text{H,th}}=10\,\text{cm}^{-3}$, which we do predict to be exceeded in the cores of the most-massive RELHICs.
However, as noted by \citetalias{benitez-llambayPropertiesDarkLCDM2017}, the \citet{schayeStarFormationThresholds2004} prescription is strictly valid only for metallicity $Z\geq 10^{-4}\,Z_\odot$, and diverges for lower metallicities.

The value of $n_{\text{H,th}}$ predicted for RELHICs is a somewhat arbitrary value imposed to avoid this behaviour.
Consequently, a rigorous investigation of the conditions under which a molecular phase may form in pristine gas would require a self-consistent treatment of the relevant atomic processes in our radiative transfer code, which we do not attempt to implement here.
As detailed in \S 2.3 of \citetalias{sykesFluorescentRingsStarfree2019}, we have instead considered $\Hmol$ formation as a post-processing step, finding that our adopted column density threshold of $N_{\HI,\,\text{max}}=10^{23}\,\text{cm}^{-2}$ corresponds to the threshold above which formation of $\Hmol$ occurs.
Additionally, the upper bound on halo mass of $\Mv\leq10^{9.6}\,\Msun$ that this threshold implies is consistent with the masses found for the largest RELHICs in \textsc{Apostle}.
While more detailed modelling may result in refinements to our predictions, we expect the existence of a window in halo mass for which predominantly `dark' haloes may contain optically-thick gas to be robust to these changes.
Furthermore, we have shown that provided this assumption holds, the results presented here are insensitive to the precise column density threshold (and hence mass scale) chosen.

We must also address the fact that RELHICs are an entirely theoretical prediction, and discuss the prospects for their detection via observations.
RELHICs exhibiting the brightest fluorescent rings are expected to be intrinsically rare, due to the narrow range of halo masses these objects may have.
In \citetalias{sykesFluorescentRingsStarfree2019}, we used \textsc{Apostle} to obtain a predicted count of $3^{+2.6}_{-2.0}$ RELHICs with $\Sigma_{\Ha,\text{max}}>10^{-20}\,\sbunit$ and a projected ring diameter ${\geq}1\,$kpc located within a 3\,Mpc volume centred on the Milky Way.
This rarity, coupled with the fact that even the brightest emission from RELHICs is still very faint for current technology, means detecting them is challenging at present.
As such, a blind $\Ha$ survey using current instrumentation (e.g. the MUSE instrument at the VLT) is likely unfeasible, requiring several tens of hours of integration time per field.
However, there remain reasons for optimism.
By nature, RELHICs contain substantial reservoirs of neutral hydrogen, making them bright $\HI$ 21cm emitters.
They are therefore expected to appear in existing deep $\HI$ surveys, such as \textsc{Alfalfa} \citep{giovanelliAreciboLegacyFast2005} and \textsc{Halogas} \citep{healdWesterborkHydrogenAccretion2011}.
In particular, the catalogue of ultra-compact high velocity clouds identified in \textsc{Alfalfa} \citep{adamsCatalogUltracompactHigh2013} have properties consistent with the expected $\HI$ morphology of the most massive RELHICs, as demonstrated by \citet{benitez-llambayPropertiesDarkLCDM2017}.
Furthermore, planned surveys with the Square Kilometre Array and its precursors \citep[e.g. \textsc{MeerKAT};][]{deblokOverviewMHONGOOSESurvey2018} will permit detection of $\HI$ sources with column densities down to $N_{\HI}\sim10^{16}\,\pcm$~\citep{poppingObservationsIntergalacticMedium2015,powerGalaxyFormationDark2015}.
This level of sensitivity is sufficient to yield $\HI$ detections of all but the lowest-mass RELHICs in the Local Group.

The $\HI$ catalogues produced by these surveys may be used in conjunction with deep broad-band imaging to identify 21cm sources with no associated stellar continuum as promising targets.
Ultra-deep pointed observations or stacking analysis of objects selected in this way have the potential to reveal the presence of one or more RELHICs, and could be used to obtain the measurements necessary to apply the techniques we have discussed here.
Thus, RELHICs remain a promising target for further study.
Their successful detection would not only constitute an additional verification of the prevailing CDM cosmological model, but as we have shown in this work, would also provide new insight into properties as disparate as the composition of the primordial Universe and the low-redshift intergalactic radiation environment of the Local Group.

\vspace{-\baselineskip}
\section*{Acknowledgements}

CS acknowledges support by a Science and Technology Facilities Council (STFC) studentship [grant number ST/R504725/1].
MF \& TT acknowledge support by the STFC [grant number ST/P000541/1]. This project has received funding from the European Research Council (ERC) under the European Union's Horizon 2020 research and innovation programme (grant agreement No. 757535).
During this work, RJC was supported by a Royal Society University Research Fellowship.
This work used the DiRAC@Durham facility managed by the Institute for Computational Cosmology on behalf of the STFC DiRAC HPC Facility (www.dirac.ac.uk). The equipment was funded by BEIS capital funding via STFC capital grants ST/K00042X/1, ST/P002293/1, ST/R002371/1 and ST/S002502/1, Durham University and STFC operations grant ST/R000832/1. DiRAC is part of the National e-Infrastructure. This work has benefited from the public Python packages \textsc{NumPy}, \textsc{SciPy}, \textsc{AstroPy} and \textsc{Matplotlib}.



\bibliographystyle{mnras}
\bibliography{He_letter}








\bsp	
\label{lastpage}
\end{document}